\documentclass[pre,preprint,floatfix]{revtex4}
\usepackage{amssymb,mathptm,multirow}
\usepackage{amsmath,amssymb,amsfonts,bm}
\usepackage{graphicx,epsfig,color}
\definecolor{orange}{rgb}{1.0,0.7,0.0}
\begin{document}
\newcommand{\be}{\begin{equation}}
\newcommand{\ee}{\end{equation}}
\title{A box-covering algorithm for fractal scaling in scale-free networks}
\author{J.S. Kim,$^1$ K.-I. Goh,$^{2,3}$ B. Kahng,$^{1,2}$ and D. Kim$^1$}
\affiliation{\\
{$^1$ CTP {\rm \&} FPRD, School of Physics and Astronomy, Seoul
National University, NS50, Seoul 151-747, Korea}\\
{$^2$ Center for Complex Network Research and Department of Physics,
University of Notre Dame, Notre Dame,
IN 46556, USA}\\
{$^3$ Department of Physics, Korea University, Seoul 136-713,
Korea}}
\date{\today}
\begin{abstract}
A random sequential box-covering algorithm recently introduced to
measure the fractal dimension in scale-free networks is
investigated. The algorithm contains Monte Carlo sequential steps of
choosing the position of the center of each box, and thereby,
vertices in preassigned boxes can divide subsequent boxes into more
than one pieces, but divided boxes are counted once. We find that
such box-split allowance in the algorithm is a crucial ingredient
necessary to obtain the fractal scaling for fractal networks;
however, it is inessential for regular lattice and conventional
fractal objects embedded in the Euclidean space. Next the algorithm
is viewed from the cluster-growing perspective that boxes are
allowed to overlap and thereby, vertices can belong to more than one
box. Then, the number of distinct boxes a vertex belongs to is
distributed in a heterogeneous manner for SF fractal networks, while
it is of Poisson-type for the conventional fractal objects.
\end{abstract}
\maketitle

{\bf A box-covering method is a basic tool to measure the fractal
dimension of conventional fractal objects embedded in the
Euclidean space. Such a method, however, cannot be applied to
scale-free networks that exhibit an inhomogeneous degree
distribution and the small-worldness. The Euclidean metric is not
well defined in such networks. To check the fractality, a random
sequential box-covering algorithm was recently introduced. In the
algorithm, vertices within a box can be disconnected, but
connected via a different box or boxes. Here we show that such
box-split allowance is an essential ingredient to obtain the
fractal scaling in scale-free networks, while it is inessential
for the conventional fractal objects. Moreover, the algorithm is
viewed from a different perspective that boxes are allowed to
overlap instead of being split and thereby, vertices can belong
to more than one box. Then, the number of distinct boxes a vertex
belongs to is distributed in a heterogeneous manner for
scale-free fractal networks, while it is of Poisson-type for the
conventional fractal objects.}

\section{Introduction}
\noindent Fractal objects that are embedded in the Euclidean
space have been observed in diverse phenomena~\cite{feder}. They
contain self-similar structures within them, which are
characterized in terms of non-integer dimension, i.e., the
fractal dimension $d_B$, defined in the fractal scaling relation,
\be N_B(\ell_B)\sim \ell_B^{-d_B}. \label{fractal} \ee Here
$N_B(\ell_B)$ is the minimum number of boxes needed to tile a
given fractal object with boxes of lateral size $\ell_B$. This
counting method is called the box-covering method.

Fractal scaling (\ref{fractal}) was also observed recently~\cite{ss}
in real-world scale-free (SF) networks such as the world-wide
web~\cite{www}, metabolic network of {\em Escherichia coli} and
other microorganisms \cite{metabolic}, and protein interaction
network of {\it Homo sapiens}~\cite{dip}. SF networks~\cite{ba} are
those that exhibit a power-law degree distribution $P_d(k)\sim
k^{-\gamma}$. Degree $k$ is the number of edges connected to a given
vertex. For such fractal networks, since their embedded space is not
Euclidean, the Euclidean metric is replaced by the chemical
distance.

One may define the fractal dimension in another manner through the
mass-radius relation. The average number of vertices $\langle
M_C(\ell_C) \rangle$ within a box of lateral size $\ell_C$, called
average box mass, scales in a power-law form, \be \langle
M_C(\ell_C)\rangle \sim \ell_C^{d_B}, \label{boxcovering}\ee with
the fractal dimension $d_B$. This counting method is called the
cluster-growing method below. Hereafter, the subscripts $B$ and $C$
represent the box-covering and the cluster-growing methods,
respectively. The formulas (\ref{fractal}) and (\ref{boxcovering})
are equivalent when the relation $N \sim N_B(\ell_B)\langle
M_C(\ell_C) \rangle$ holds for $\ell_B=\ell_C$. Such case can be
seen when fractal objects are embedded in the Euclidean space.
However, for SF fractal networks, the relation (\ref{boxcovering})
is replaced with the small-world behavior, \be \langle M_C(\ell_C)
\rangle \sim e^{\ell_C/\ell_0}, \label{sw} \ee where $\ell_0$ is a
constant. Thus, the fractal scaling can be found in the box-covering
method, but not in the cluster-growing method for SF fractal
networks.

To understand this seemingly contradictory relations, here we
investigate generic nature of the box-covering method in SF networks
in comparison of the cluster-growing method. Owing to the
inhomogeneity of degrees in SF fractal networks, the way of covering
a network can depend on detailed rules of box-covering methods.
Recently, a new box-covering algorithm was introduced by the current
authors~\cite{goh2006,jskim}. In fact, this algorithm shares a
common spirit with the one previously introduced by Song {\it et
al.}~\cite{ss}, however, details differ from one another in the
following perspective: Our algorithm, called random sequential (RS)
box-covering method, contains a random process of selecting the
position of the center of each box. A new box can overlap preceding
boxes. In this case, vertices in preassigned boxes are excluded in
the new box, and thereby, vertices in the new box can be
disconnected within the box, but connected through a vertex (or
vertices) in a preceding box (or boxes). Nevertheless, such a
divided box is counted as a single one. Detailed rule is described
in the next section. Such counting method is an essential ingredient
to obtain the fractal scaling in fractal networks; whereas, it is
inessential for regular lattice and conventional fractal objects
embedded in the Euclidean space.

Next, we count how many boxes a vertex belongs to in the
cluster-growing algorithm, where boxes are allowed to overlap. For
the SF fractal network, the fraction of vertices counted $f$ times
decays with respect to $f$ in a nontrivial manner, while for the
square lattice and a conventional fractal object, it decays in a
Poisson-type manner. We note that the Sierpinski gasket is used here
as a fractal object embedded in the Euclidean space. Such distinct
features arising in the SF fractal networks enables the coexistence
of the two contradictory notions of the fractality and
small-worldness.

\section{Random sequential box-covering}

Here we describe a new box-covering method, which takes steps as
follows: We start with all vertices labeled as {\em not burned}.
Then,
\begin{enumerate}
\item[(i)] Select a vertex randomly at each step; this vertex serves
as a seed.
\item[(ii)] Search the network by distance $\ell_B$ from the seed
and burned all vertices found but not burned yet. Assign {\em newly
burned vertices} to the new box. If no newly burned vertex is found,
the box is discarded.
\item[(iii)] Repeat {(i)} and {(ii)} until all vertices
are assigned to their respective boxes.
\end{enumerate}

\begin{figure}[t]
\centerline{\epsfxsize=5cm \epsfbox{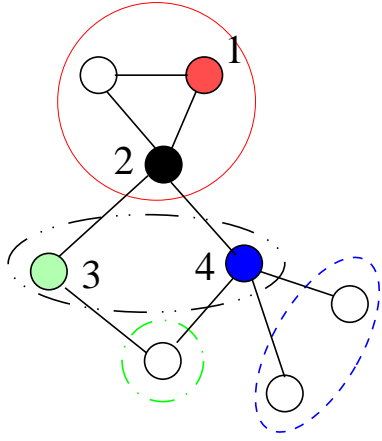}} \caption{(Color
online) Schematic illustration of the RS box-covering algorithm
introduced~\cite{goh2006,jskim}. Vertices are selected randomly, for
example, from vertex 1 to 4 successively. Vertices within distance
$\ell_B=1$ from vertex 1 are assigned to a box represented by solid
(red) circle. Vertices from vertex 2, not yet assigned to their
respective box are represented by dashed-doted-doted (black) closed
curve, vertices from vertex 3 are represented by dashed-doted
(green) circle and vertices from vertex 4 are represented by dashed
(blue) ellipse.} \label{box_method}
\end{figure}

The above method is schematically illustrated in
Fig.~\ref{box_method}. A different Monte Carlo realization of this
procedure ((i)--(iii)) may yield a different number of boxes for
covering the network. In this study, for simplicity, we choose the
smallest number of boxes among all the trials. To obtain the
power-law behavior of the fractal scaling, we needed at most
${\mathcal O}(10)$ Monte Carlo trials for all fractal networks we
study. It should be noted that the box number $N_B$ we employ is not
the minimum number among {\em all} the possible tiling
configurations. Finding the actual minimum number over all
configurations is a challenging task, which could not be reached by
the Monte Carlo method.

\begin{figure}
\centerline{\epsfxsize=7cm \epsfbox{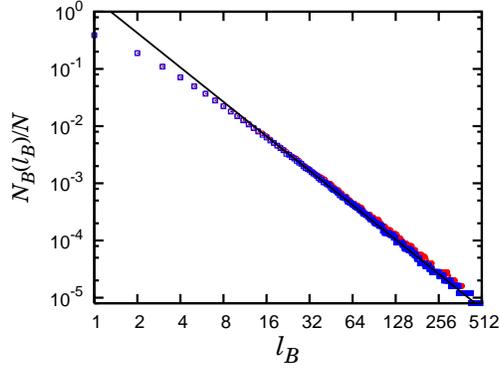}} \caption{(Color
online) Fractal scaling analysis for the two dimensional square
lattice with the box-covering algorithm. Shown are the result of one
Monte Carlo trial (\textcolor{red}{$\circ$}) and that obtained from
20 Monte Carlo trials (\textcolor{blue}{$\square$}). From the
least-square-fit of the data (straight line), the fractal dimension
is measured to be $\approx$2.0 as expected.}\label{twodim}
\end{figure}

To check the validity of our algorithm, we first apply our method to
the two dimensional regular lattice in Fig.~\ref{twodim}. While our
method may perform inefficiently in the highly regular structure due
to the step in which already box-assigned vertices can be selected
as seeds, taking about a few hours of cpu time for system size
$N=500 \times 500$, we find that our method can still yield the
correct dimension $\approx$2.0 for the two dimensional square
lattice, as shown in Fig.~\ref{twodim}. We also show that the number
of Monte Carlo trials is not crucial to obtain the fractal scaling.

\begin{figure}[t]
\centerline{\epsfxsize=9cm \epsfbox{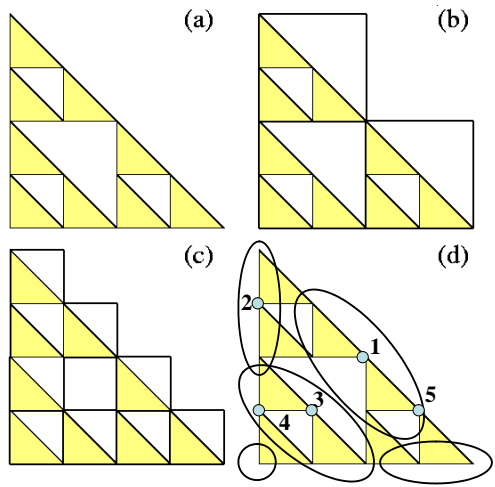}} \caption{(Color
online) A fractal object, the Sierpinski gasket with the second
generation (a). Conventional box covering based on the Euclidean
metric with size $\ell_E=2$ (b) and $\ell_E=1$ (c). Box covering
based on chemical distance $\ell_B=1$ by using the RS box-covering
method (d). Seed vertices $1\to 5$ are selected successively.}
\label{sierpinski}
\end{figure}
\begin{figure}
\centerline{\epsfxsize=7cm \epsfbox{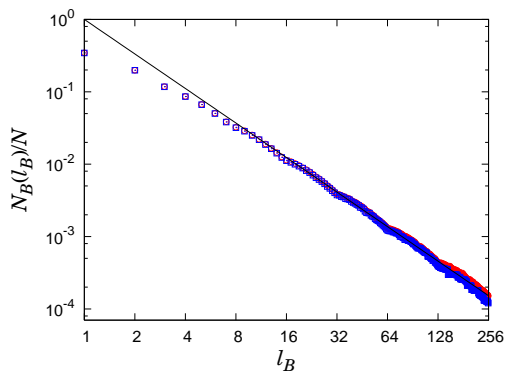}} \caption{(Color
online)Fractal dimension of the Sierpinski gasket with the 12th
generation measured by using the RS box-covering method. Solid line
is guideline with slope $-\ln 3/\ln 2$. Shown are the result of one
Monte Carlo trial (\textcolor{red}{$\circ$}) and that obtained from
20 Monte Carlo trials (\textcolor{blue}{$\square$}).}
\label{sier_frac_dim}
\end{figure}

Fig.\ref{sierpinski} shows the box-covering method applied to the
Sierpinski gasket with the third generation (a). One can find easily
that $N_B(\ell_E=2)=3$ in (b) and $N_B(\ell_E=1)=9$ in (c) when
lateral size $\ell_E$ is taken as the conventional Euclidean metric.
The obtained $N_B(\ell_E)$ are the minimum numbers of boxes needed
to tile the object for each case. In Fig.\ref{sierpinski}(d), we
show a configuration in box-covering ensemble obtained from our
current algorithm with distance $\ell_B=1$. One can see that
$N_B(\ell_B)$ can vary depending on Monte Carlo trials. We show,
however, that the fractal dimension is obtained by using the
conventional method in (b) and (c), $d_B=-\ln 3/\ln 2$. Numerical
value is obtained from the Sierpinski gasket with the 12th
generation, composed of 265,721 vertices, and the fractal scaling is
shown in Fig.~\ref{sier_frac_dim}.

\begin{figure}
\centerline{\epsfxsize=7cm \epsfbox{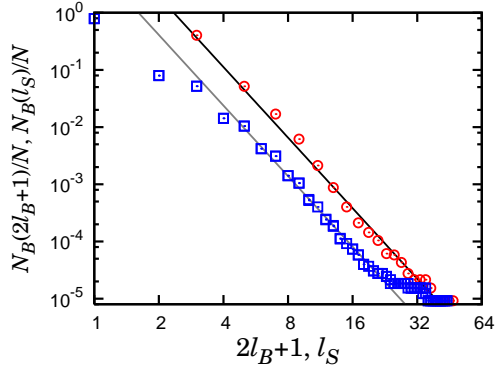}} \caption{ (Color
online) Comparison of the box-covering methods introduced by Song
{\it et al.}~\cite{ss} (\textcolor{blue}{$\square$}) and in this
paper (\textcolor{red}{$\circ$}) for the world-wide web. The results
obtained from the two box-covering methods applied to the world-wide
web are plotted here. The two methods yield the same fractal
dimension $d_B\approx 4.1$. The method introduced by Song {\it et
al.} is more optimal than ours in the viewpoint that $N_B(\ell_S) <
N_B(2\ell_B+1)$. } \label{frac}
\end{figure}

Our algorithm also generates the same fractal dimensions for SF
fractal networks such as the world-wide web, the metabolic network
of {\it E. coli}, the protein interaction networks of {\it H.
sapiens} and {\it S. cerevisiae} as obtained by Song {\em et
al.}~\cite{ss}. Fig.\ref{frac} shows the fractal scaling for the
world-wide web, displaying the same fractal dimension $d_B\approx
4.1$. In comparison of the method by Song {\em et al.}, ours is
easier to implement, because it does not contain the procedure for
constraining the maximum separation within a box and is carried out
in random sequential manner in box covering.

\begin{figure}
\centerline{\epsfxsize=7cm \epsfbox{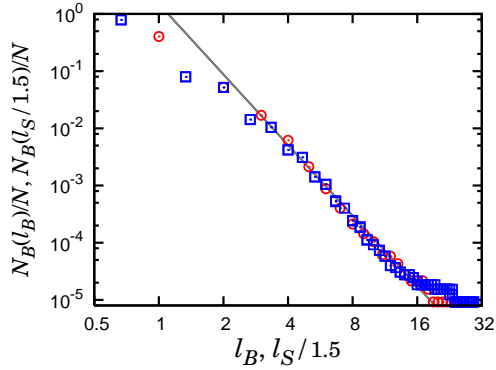}} \caption{(Color
online) Fractal scaling analysis for the world-wide web by the two
box-covering algorithms, that of Song {\it et al.}\
(\textcolor{blue}{$\square$}) and of ours
(\textcolor{red}{$\circ$}). For comparison, the horizontal scale for
that of Song {\em et al.}\ is rescaled as $\ell_S/1.5 \to \ell_B$,
by which we get the overlap of two curves obtained from the
different algorithms. }\label{rescaling}
\end{figure}

The particular definition of box size has proved to be inessential
for fractal scaling. It is rather inappropriate to compare the
length scale $\ell_B$ with that $\ell_S$ used in Ref.~\cite{ss},
because the two methods involve different definitions of the boxes.
What is interesting, however, is that there exists a linear
relationship, for example $\ell_S/1.5 \to \ell_B$ in the case of the
world-wide web as shown in Fig.~\ref{rescaling}. This linear
relationship indicates that despite the difference in the two
algorithms, such a difference does not lead to qualitatively
different fractal-scaling behaviors.

\section{Overlap of box covering and vertices disconnectedness}

\begin{figure}
\includegraphics[width=7cm]{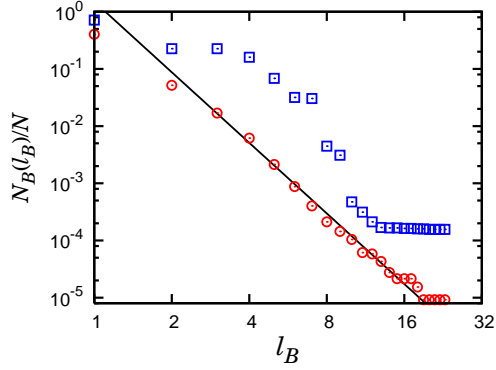}
\caption{(Color online) Fractal scaling analysis for the world-wide
web with the RS box-covering algorithm (\textcolor{red}{$\circ$})
and its variant that disallows disconnected boxes
(\textcolor{blue}{$\square$})} \label{disconnect}
\end{figure}

\begin{figure}
\includegraphics[width=7cm]{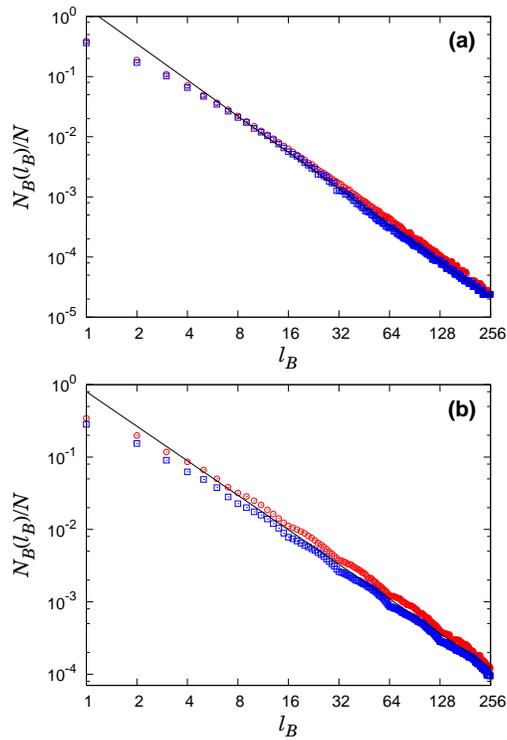}
\caption{(Color online) Fractal scaling analysis for the square
lattice (a) and the Sierpinski gasket (b) with the RS box-covering
algorithm (\textcolor{red}{$\circ$}) and its variant that disallows
disconnected boxes (\textcolor{blue}{$\square$}). Solid lines are
guidelines with slopes of $-2$ in (a) and $-\ln 3/\ln 2$ in (b).}
\label{disconn_regular}
\end{figure}

It is interesting to note that in our algorithm vertices can be
disconnected within a box, but connected through a vertex (or
vertices) in a different box (or boxes) as in the case of box 2
shown in Fig.~\ref{box_method}. On the other hand, if we construct a
box with only connected vertices, for example, box 2 is regarded as
two separate boxes, then the power-law behavior Eq.~(\ref{fractal})
is not observed for the world-wide web as shown in
Fig.~\ref{disconnect}. We check if such difference appears even for
a regular lattice and a fractal object embedded in the Euclidean
space. Figs.~\ref{disconn_regular} show that such different behavior
does not occur for the square lattice in two dimensions and the
Sierpinski gasket. We show that such fact originates from the
inhomogeneity of degrees in SF networks as follows: Owing to their
large degree, hub vertices can be assigned to boxes earlier than
other vertices when their neighbors are selected as seeds of boxes.
Once hub vertices are assigned to one of the boxes, they can make
subsequent boxes disconnected when vertices in those boxes are
connected via hub vertices. Box 2 in Fig.~\ref{box_method} is such a
case. In SF networks, such cases occur at a non-negligible rate.

\begin{figure}
\centerline{\epsfxsize=7cm \epsfbox{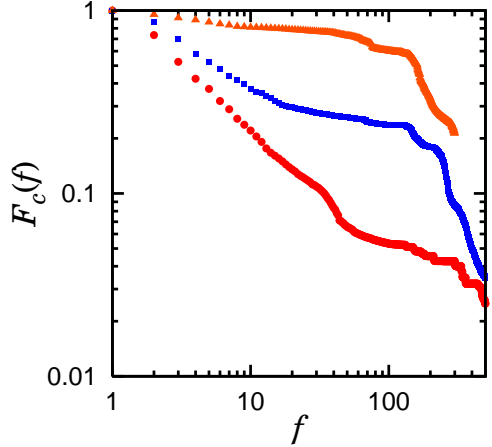}} \caption{(Color
online) Cumulative fraction $F_c(f)$ of the vertices counted $f$
times in the cluster-growing algorithm. $F_c(f)$ follows a power law
in the small $f$ region, where the slope depends on box size
$\ell_C$. However, for large values of $f$, the data largely deviate
from the value extrapolated from the power-law behavior. Data are
presented for $\ell_C=2$ (\textcolor{red}{$\bullet$}), $\ell_C=3$
(\textcolor{blue}{$\blacksquare$}), and $\ell_C=5$
(\textcolor{orange}{$\blacktriangle$}).} \label{histogram}
\end{figure}

\begin{figure}
\centerline{\epsfxsize=7cm \epsfbox{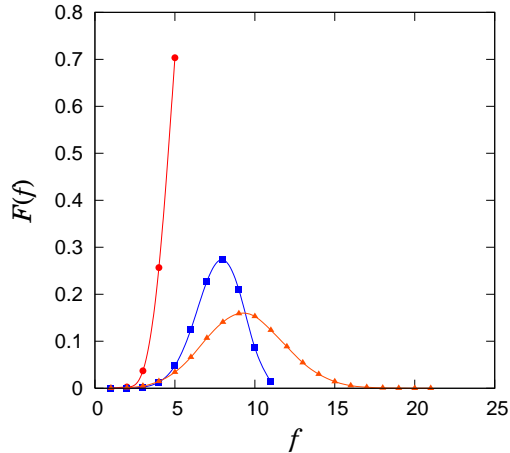}} \caption{(Color
online) Fraction $F(f)$ of the vertices counted $f$ times in the
cluster-growing algorithm for the Sierpinski gasket with the 12th
generation composed of 265,721 vertices. $F(f)$ follows a
Poisson-type distribution. Data are presented for $\ell_C=2$
(\textcolor{red}{$\bullet$}), $\ell_C=3$
(\textcolor{blue}{$\blacksquare$}), and $\ell_C=5$
(\textcolor{orange}{$\blacktriangle$}).} \label{histogram_SG}
\end{figure}

To study the fraction of disconnected boxes quantitatively, we
invoke the cluster-growing approach. In this approach, boxes are
allowed to overlap, and thereby, a vertex can belong to more than
one box. Thus, the extent of overlap of the boxes during the tiling
can provide important information on the fraction of disconnected
boxes in the box-covering method. In this regard, we reported the
cumulative fraction $F_c(f)$ of vertices counted $f$ times or more
in the cluster-growing method for the world-wide web in \cite{jskim}
and is reproduced in Fig.~\ref{histogram}. The cumulative fraction
$F_c(f)$ is likely to follow a power law for small $f$, thereby
indicating that the overlaps occur in a non-negligible frequency
even for a small distance $\ell_C$. The associated exponent
decreases with increase in box size $\ell_C$ as the chances of
overlaps increase. However, for large values of $f$, the large
fraction of vertices counted exceed the frequency extrapolated from
the power-law behavior. For the square lattice and the Sierpinski
gasket, however, the fraction $F(f)$ follows a bounded distribution
with a peak at small $f$ as shown in Fig.~\ref{histogram_SG}. Thus,
for the fractal networks like the world-wide web, there are a
significant number of vertices that are counted quite a few times in
the cluster-growing method, but such vertices are extremely rare in
the conventional fractal objects such as the Sierpinski gasket. Such
multiple counting due to overlap is excluded in the box-covering
method. This exclusion effect makes the average mass of a box in the
box-covering method significantly lower than that in the
cluster-growing method.


\begin{figure}
\includegraphics[width=7cm]{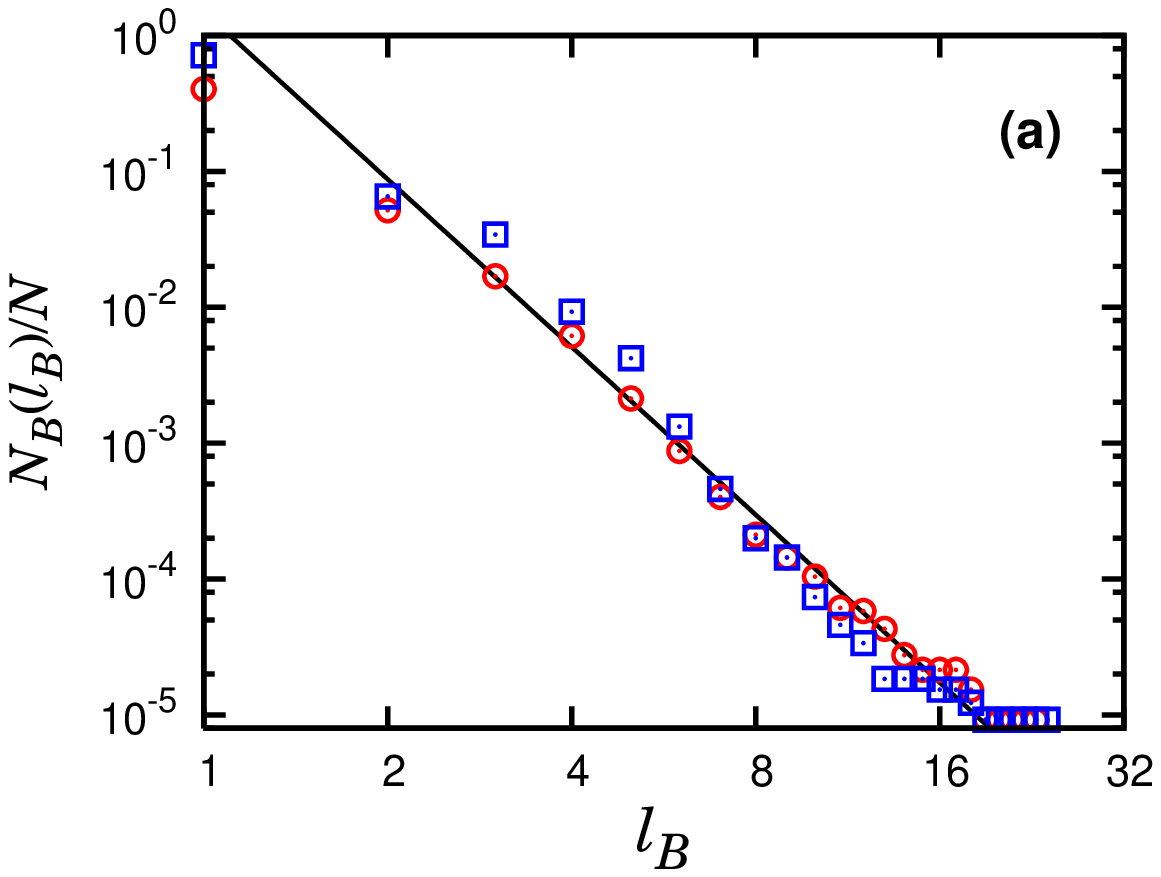}
\includegraphics[width=7cm]{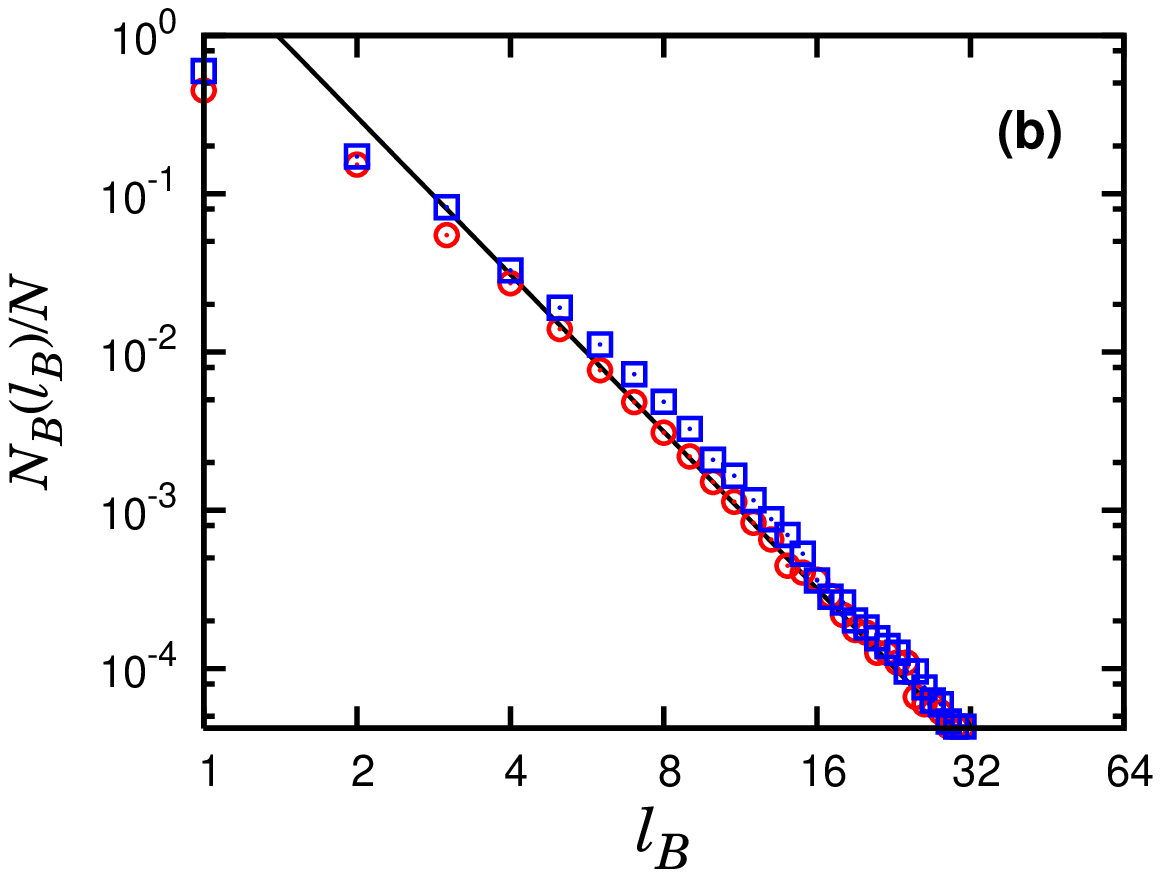}
\caption{(Color online) Fractal scaling analysis with the rules that
allow (\textcolor{red}{$\circ$}) or disallow
(\textcolor{blue}{$\square$}) the already-box-burned vertices to be
chosen as the roots of new boxes for the world-wide web (a), and the
fractal model (b).}\label{efficiency}
\end{figure}

Next, one may wonder if the RS box-covering algorithm can be
improved in efficiency by excluding already-burned vertices from the
list of the root candidates of new boxes. In Fig.~\ref{efficiency},
we compare the fractal scaling behaviors obtained from the two cases
of keeping or excluding already-burned vertices from the list for
two networks: the world-wide web (a) and the fractal model
introduced in~\cite{goh2006}. We find that the two cases exhibit
somewhat different behaviors. If the already-burned vertices are
excluded from the next selection, the power-law behavior is not
obtained for the world-wide web. However, they exhibit similar
power-law behaviors for the fractal model, even though the two data
sets show somewhat deviations.

\section{Conclusions and Discussion}
We have studied various features of the random sequential
box-covering algorithm by applying it to a SF fractal network, the
world-wide web, and a regular and a conventional fractal object, the
square lattice and the Sierpinski gasket, respectively. Results
obtained from the two classes of networks exhibit distinct feature.
The condition that vertices in a box can be disconnected in the
box-covering method turns out to be an essential ingredient to have
the fractal scaling for a SF fractal network, however, it is
irrelevant for a regular lattice and a conventional fractal object
embedded in the Euclidean space. We also found that the fraction of
vertices counted $f$ times in the cluster-growing method exhibits a
non-trivial behavior for the former, while it does a trivial
behavior for the latter. The two results are complementary; thereby,
the SF fractal network exhibits the fractal scaling (\ref{fractal})
in the box-covering and the small-world behavior (\ref{sw}) in the
cluster-growing method. Finally, it is noteworthy that our
box-covering algorithm is a modification of the algorithm
used in the random sequential packing problem~\cite{packing}.\\

This work was supported by KRF Grant No.~R14-2002-059-010000-0 of
the ABRL program funded by the Korean government (MOEHRD). Notre
Dame's Center for Complex Networks kindly acknowledges the support
of the National Science Foundation under Grant No. ITR DMR-0426737.

\end{document}